\documentclass[12pt]{article}
\usepackage{amsfonts,amssymb,epsfig,amsmath,times}
\usepackage{float}
\restylefloat{table}
\renewcommand{\text}[1]{#1}

\newcommand{\be}{\begin{equation}}
\newcommand{\ee}{\end{equation}}
\newcommand{\ben}{\begin{displaymath}}
\newcommand{\een}{\end{displaymath}}
\newcommand{\bea}{\begin{eqnarray}}
\newcommand{\eea}{\end{eqnarray}}
\newcommand{\bean}{\begin{eqnarray*}}
\newcommand{\eean}{\end{eqnarray*}}
\newcommand{\baa}{\begin{array}}
\newcommand{\eaa}{\end{array}}
\newcommand{\nn}{\nonumber \\}
\newcommand{\ba}{\begin{array}}
\newcommand{\ea}{\end{array}}
\newcommand{\bi}{\begin{itemize}}
\newcommand{\ei}{\end{itemize}}

\def\g{\gamma}

\def\G{\Gamma}
\def\d{\delta}

\def\e{\epsilon}

\def\otaula{\begin{tabular}}
\def\ctaula{\end{tabular}}

\def\bnum{\begin{enumerate}}
\def\enum{\end{enumerate}}

\def\CR{\mathcal{R}}
\def\CM{\mathcal{M}}

\def\8M{$\CM_8$}

\def\be{\begin{equation}}
\def\ee{\end{equation}}
\def\G{\Gamma}
\def\g{\gamma}
\def\ei{e^{\underline{i}}}

\def\e1{e^{\underline{1}}}
\def\1u{\underline{1}}
\def\2u{\underline{2}}

\def\0u{\underline{0}}
\def\e{\epsilon}
\def\target{$\CR^{1,1}\times \mathcal{M}_8$ }
\def\target2{$\CR^{1,1}\times \mathcal{M}_8$,}
\def\9G{\G_{\underline{9}}}

\def\1f{f_1^{1/2}}
\def\2f{f_2^{1/2}}
\def\4f{f_4^{1/2}}

\begin{document}

\makeatletter
\renewcommand{\theequation}{\thesection.\arabic{equation}}
\@addtoreset{equation}{section}
\makeatother

\baselineskip 18pt

\begin{titlepage}

\vfill

\begin{flushright}
KIAS-P08069
\end{flushright}

\vfill

\begin{center}
   \baselineskip=16pt
   {\Large\bf BPS D0-D6 Branes in Supergravity}
   \vskip 2cm
      Ki-Myeong Lee $^{1}$, Eoin \'{O} Colg\'{a}in $^{1}$, Hossein Yavartanoo$^{1}$ and K. P. Yogendran$^{2}$
         \vskip .6cm
      \begin{small}
      $^1$\textit{Korean Institute for Advanced Study, \\
        Seoul 130-722, Korea}
        \end{small}\\*[.6cm]
      \begin{small}
      $^2$\textit{ Helsinki Institute of Physics,\\ University of Helsinki,  
Finland}
        \end{small}\\*[.6cm]
\end{center}

\vfill
\begin{center}
\textbf{Abstract}
\end{center}

\begin{quote}
We analyse 1/2 BPS IIA Dp-brane supergravity solutions with $B$-fields and their Killing spinor equations. Via probe analysis, we rederive the supersymmetry conditions for D0-Dp with $B$-fields. In the case of D6 with $B$-fields, the D0-probe sees a multi-centred BPS configuration where the $B$-fields give the location of a wall of marginal stability. Finally we go beyond the probe approximation and construct a 1/8 BPS supergravity solution for a fully back-reacted D0-D6 with $B$-fields. 
\end{quote}

\vfill

\end{titlepage}
\section{Introduction}
Once the connection to dual non-commutative gauge-theories was realised \cite{Connes:1997cr}, much work was done analysing the supersymmetry conditions for D0-Dp brane systems in the presence of $B$-fields \cite{Witten:2000mf,Seiberg:1999vs}. In short, D0-D2 will not be supersymmetric for any finite $B$-field, D0-D4 is 1/4 BPS with (anti-)self-dual fields $B_{12} = \pm B_{34}$, and D0-D6 requires $\pm B_{12} B_{34} \pm B_{34} B_{56} \pm B_{12} B_{56} = \pm 1$ for it to be 1/8 BPS. As T-duality maps $B$-fields to frame rotations, these systems are T-dual to rotated brane configurations whose supersymmetry conditions have also appeared in the literature \cite{Berkooz:1996km}.  

Once the supersymmetry conditions have been identified, one of the subsequent steps is identifying supergravity solutions. 
In the spirit of the AdS/CFT, the supergravity duals of non-commutative field theories were studied in \cite{Maldacena:1999mh} by considering branes on tilted tori. Unless other branes are present, despite the obfuscating zoo of induced fluxes, these Dp-brane $B$-field solutions are 1/2 BPS and there are no constraints on the $B$-fields. Apart from the Dp-brane with $B$-fields solution, the only other supersymmetric solution appearing explicitly in the IIA literature is that of  D0-D4 $B_{12}, B_{34}$ \cite{Lifschytz:1996ng}. To the extent of our knowledge, the supersymmetric ten-dimensional D0-D6 with $B$-fields solution has not been written down explicitly. However, in both four and five-dimensions, prescriptions have been given for constructing such solutions.  

In the absence of fully back-reacted solutions, there are various techniques to glean a better understanding of the physics. 
In this paper, we make use of both string theory scattering amplitudes and DBI D0-probe potentials to get a better picture. 
From analysing scattering amplitudes, we see evidence for three regimes: one with sub-critical $B$-fields where D0 is repelled from D6; a critical $B$-field regime where there is no force; and a super-critical $B$-field where the D0 is attracted. So, for large enough $B$-fields one could imagine a scenario where D0 sits on top of D6. However, this is not the whole story. When we consider the back-reaction of the D6 with $B$-field solution, we find that a D0 ``sees'' a potential. For the critical $B$-field, the minimum is at infinity, while as $B$ increases, the minimum migrates inwards towards the D6, but never reaches the D6-brane for any finite $B$-field. All of this chimes well with the work of Denef and Moore \cite{Denef:2000nb, Denef:2007vg}. 

In the literature, in lower dimensions, there are works allowing descriptions of a back-reacted D0-D6 BPS state.  In the elaborate and far-reaching works of Denef and Moore, a picture emerges in four-dimensions, of BPS bound states of D-branes either at a single point or as multi-centred composites \cite{Denef:2000nb}. In particular for the D0-D6 system, in the presence of large enough $B$-fields, a wall of marginal stability exists separating a bound D0-D6 state at finite separation from its infinitely separated constituents \cite{Denef:2007vg}. 

Via the ``4D-5D" connection \cite{Bena:2005ay}, these composites are related to a whole host of supersymmetric super-tube, black ring and black hole solutions in five-dimensions \cite{Gauntlett:2003fk}. For a review of these solutions, see \cite{Emparan:2006mm}. With these solutions in hand, their microstate counting is an area that has received much attention. In particular, for non-supersymmetric D0-D6, a description of microstates in terms of intersecting D3-branes was proposed recently in \cite{Emparan:2006it}. 

While we were winding up this project, we became aware of \cite{Emparan} which was also nearing completion, and overlaps with some of the material in this paper. 

The structure of the rest of this paper runs as follows.  
In section 2, we write down 1/2 BPS D2, D4 and D6-branes with background $B$-fields in IIA supergravity. We solve the Killing spinor equations, and compare the projectors in each case with a Dp-brane $\kappa$-symmetry probe. In section 3, as a stepping stone towards a back-reacted description, we introduce a D0-probe and determine the amount of preserved supersymmetry again via $\kappa$-symmetry. We then determine the potential seen by a static D0-probe as a function of the $B$-fields. In the case of D6, we recover further evidence for the multi-centres of Denef-Moore. We also perform a complementary calculation in string theory, and compare the results with the DBI potentials. In section 4, we explicitly construct a fully back-reacted D0-D6 solution with $B$-fields which preserves 1/8 supersymmetry. This solution is not a black hole, but we show that the addition of some extra charges will produce a black hole at the location of the D6. Our conventions and anything that deviates from the thrust of the main text appears in the appendix.

\section{Dp-branes with $B$-fields}
Dp-brane solutions with $B$-fields in IIA supergravity may be easily constructed by tilting tori and T-dualising (or alternatively performing an $O(d,d)$ transformation on isometric directions). Either way, the addition of a $B$-field, mimics a rotation of the brane configuration. By examining the Killing spinor equations for the resulting solutions, one may see that the overall effect of turning on a $B$-field in a Dp-brane background, is simply to rotate the Killing spinor. As such no supersymmetry is broken and all the solutions we present in this section will be 1/2 BPS. 

$B$-fields were added to Dp-branes in \cite{Maldacena:1999mh}, where the backgrounds were used to explore the duals of non-commutative field theories. Here we review the D2 case. The approach is to start with D1 wrapped on one cycle of a $T^{2}$, $x_{1}$, while at the same time being smeared over the other cycle, $x_{2}$. 
We now tilt this torus by performing an area-preserving coordinate transformation
before T-dualising on the new $x_{2}$ direction. The final solution is 
\bea
\label{D2B}
ds^{2}_{str} &=& f_{2}^{-1/2}[-dx^{2}_{0} + h(dx^{2}_{1}+dx^{2}_{2})] +f_{2}^{1/2} (dr^{2} + r^{2} d \Omega_{6}^{2}), \nn
e^{\phi} &=& f_{2}^{1/4} h^{1/2}, 
\mbox{\hspace{5mm}} 
 B_{12} = \frac{\sin \theta}{\cos \theta} f_{2}^{-1} h, \nn
F_{r0} &=& \sin \theta \partial_{r} f_{2}^{-1}, 
 \mbox{\hspace{5mm}} 
F_{r012} = - \cos \theta h \partial_{r} f_{2}^{-1}, 
\eea
where
\be
\label{hf}
f_{2} = 1 + \frac{Q_{2}}{r^5}, \quad h^{-1} = \sin^{2} \theta f_{2}^{-1} + \cos^{2} \theta.  
\ee
As a quick check, note that righting the torus by taking the $\theta \rightarrow 0$ limit, we find, as expected, the D2 brane solution without $B$-field \footnote{Throughtout this text $2 \pi \alpha' = 1$, so $B_{2i-1\;2i} = b_{i}$.}. 

\subsection{Rotated Killing spinors}
In this section we explore the effect of how turning on a $B$-field in a Dp-brane background affects the Killing spinor equations. In the process, we verify that all these solutions are 1/2 BPS. For clarity we again focus on D2. 

We begin with the dilatino variation for \textit{pure} D2 
\bea
\d \lambda &=&  - \frac{1}{4} f_{2}^{3/4} \partial_{r} f_{2}^{-1} \G^{r} \tilde{\e}
- \frac{1}{4} f_{2}^{3/4} \partial_{r} f_{2}^{-1} \G^{r012} \tilde{\e}.
\eea
It can be quickly verified that $\G^{012} \tilde{\e} = - \G_{012} \tilde{\e} = - \tilde{\e}$ satisfies this equation, as expected. For a D2 with a $B$-field (\ref{D2B}) the dilatino variation may be re-written
\bea
\d \lambda^{(B)} &=& -\frac{1}{4} f_{2}^{3/4} \partial_{r} f_{2}^{-1} \G^{r} \left(1 - e^{\alpha \G_{12} \G_{11}}  \G_{012} \right) \e \nn
&-& \frac{1}{2} s f_{2}^{1/4} h^{1/2} \partial_{r} f_{2}^{-1} \G^{r12} \G_{11} \left( e^{- \alpha \G_{12} \G_{11} } - \G_{012} \right) \e, 
\eea
where we have used $ s \equiv \sin \theta, c \equiv \cos \theta$ to compress notation and have also defined a new angle $\alpha$ \be \cos \alpha = \cos \theta h^{1/2} , \quad \sin \alpha = \sin \theta f_{2}^{-1/2} h^{1/2}. \ee  
It is clear that the projector 
\bea
\label{projD2}
e^{\alpha \G_{12} \G_{11} } \G_{012} \e & =& \e, \nn
(\cos \theta h^{1/2} \G_{012} - \sin \theta f_{2}^{-1/2} h^{1/2} \G_{11} \G_{0} ) \e &=& \e, 
\eea
satisfies the dilatino variation. It also satisfies the gravitino variations, the details of which we move to the appendix to reduce clutter. Note that in (\ref{projD2}), the upper expression corresponds to the orthonormal frame, where $B_{12} = \tan \alpha$, while the lower corresponds to coordinate frame. This distinction will be important when we examine the D0-probes in the next section.  

We draw attention again to the $\theta \rightarrow 0$ limit: we obtain the projection operation for a D2-brane i.e. $\G_{012} \e = \e $. While in the opposite limit $\theta \rightarrow \pi/2$ we find a D0-projector. It may also be readily verified that the left hand side of (\ref{projD2}) squares to unity by observing that both $\G_{012}$ and $\G_{11} \G_{0}$ anti-commute and by also making use of (\ref{hf}). 
 
For the gravitino variations \footnote{The same is not true for $\d \lambda$. Possibly this is because it is a linear combination of gravitino variation from M-theory.}, by redefining the original Killing spinor $\tilde{\e}$ in terms of the Killing spinor with $B$-field $\e$,
\be
\label{etildee}
\e = \mbox{exp} (\alpha/2 \G_{12} \G_{11} ) \tilde{\e},
\ee
it is possible to write the variations of the gravitino in the presence of the $B$-field $\psi^{(B)}$ in terms of the original variation such that
\bea
\d \psi^{(B)} &=& e^{\alpha/2 \G_{12} \G_{11}} \d \psi, \quad \mbox{temporal, D2 transverse directions}, \nn
\d \psi^{(B)} &=& e^{3 \alpha/2 \G_{12} \G_{11}} \d \psi, \quad \mbox{ $B_{12}$ parallel directions}.
\eea
Further details for D2 maybe found in the appendix. Similar rotations were observed for D4 and D6 Killing spinor equations. This observation of rotated Killing spinors echoes \cite{Hassan:1999bv}, where an extensive analysis of Killing spinors in the presence of T-duality transformations is presented. 

Once we have solved the Killing spinor equations, we may use $\kappa$-symmetry \cite{Cederwall:1996ri,Bergshoeff:1996tu} as a consistency check to verify that the  projectors are correct. For a brane configuration the fraction of preserved supersymmetry is determined by the supersymmetry condition of the gravity background coupled with the following equation:
\be
(1-\G_{\kappa}) \e = 0,
\ee 
where $\e$ is the spacetime supersymmetry parameter and $\G_{\kappa}$ is a Hermitian, traceless matrix that squares to unity. Explicitly it may be expressed as
\be
\G_{\kappa} = \frac{\sqrt{g}}{\sqrt{g+\mathcal{F}}} \sum^{\infty}_{n=0} \frac{1}{2^{n} n!} \g^{\mu_1 \nu_1...\mu_n \nu_n}\mathcal{F}_{\mu_1 \nu_1} \cdots \mathcal{F}_{\mu_n \nu_n} J^{(n)}_{(p)},
\ee  
where $g$ is the induced matrix on the Dp-brane worldvolume and $\mathcal{F}$, is in the absence of $U(1)$ Born-Infeld field, up to sign, the background $B$-field pulled-back to the worldvolume of the brane.  
For IIA Dp-branes
\be
J^{(n)}_{(p)} = (\G_{11}^{n+(p-2)/2}) \frac{1}{(p+1)!\sqrt{g}} \e^{\mu_{1}...\mu_{p+1}} \g_{\mu_1...\mu_{p+1}}.  
\ee
By considering a D2-probe with worldvolume coordinates $(t, \xi_{i})$ $i=1,2$, where 
\be
\begin{array}{cc}
X^{0} = t, \mbox{\hspace{5mm}} X^{i} = \xi^{i},
\end{array}
\ee
along with setting $\mathcal{F}=B$, we get the above projector (\ref{projD2}). 

\subsection{D4 and D6 branes with $B$-fields}
The earlier construction of D2 with $B_{12}$ generalises readily to D4 with two orthogonal $B$-fields, $B_{12}$, $B_{34}$ and D6 with three orthogonal $B$-fields, $B_{12}$, $B_{34} $ and $B_{56}$. As for D2, no supersymmetry is broken: the final solutions are 1/2 BPS. The Killing spinor equations were solved and the equations of motion verified. In the case of D6 we constructed the solution by noting the structure in the Killing spinor equations and simply reading off the solution from the gravitino variation of $ \d \psi_{0}$. We present the solutions below.   

\vspace{2mm}
\noindent
\textbf{D4 with  $\mathbf{B_{12}, B_{34}}$} 
\bea
\label{D4BB}
ds^{2}_{str} &=& f_{4}^{-1/2}[-dx^{2}_{0} + h_{1}(dx^{2}_{1}+dx^{2}_{2}) + h_{2}(dx_{3}^{2} + dx_{4}^{2})] +f_{4}^{1/2} (dr^{2} + r^{2} d \Omega_{4}^{2}), \nn
 B_{12} &=& \frac{s_1}{c_1} f_{4}^{-1} h_1, 
 \mbox{\hspace{5mm}} 
 B_{34} = \frac{s_2}{c_2} f_{4}^{-1} h_2, 
 \mbox{\hspace{5mm}} f_{4} = 1 + \frac{Q_{4}}{r^{3}},
\nn
e^{ \phi} &=& f_{4}^{-1/4} (h_1 h_2)^{1/2},  \mbox{\hspace{5mm}} h_{i}^{-1} = s_{i}^{2} f_{4}^{-1} + c_{i}^{2}, \nn
F_{r0} &=&  -s_1 s_2 \partial_{r} f_{4}^{-1}, 
 \mbox{\hspace{5mm}} 
F_{r012} = h_{1} c_1 s_{2} \partial_{r} f_{4}^{-1}, \nn
F_{r034} &=& h_{2}  c_{2} s_{1} \partial_{r} f_{4}^{-1}, 
 \mbox{\hspace{5mm}} 
F_{r01234} = - h_{1} h_{2} c_{1} c_{2} \partial_{r} f_{4}^{-1},
\eea
where the Killing spinor projector is
\be
\label{projD4}
e^{\alpha_{1} \G_{12} \G_{11}} e^{\alpha_{2} \G_{34} \G_{11}} \G_{11} \G_{01234} \e = \e, 
\ee
and as before we define
\be
\begin{array}{cc} \cos \alpha_{i} = c_{i} h^{1/2}_{i}, & \sin \alpha_{i} = s_{i} f_{4}^{-1/2} h_{i}^{1/2}\\ \end{array}.    
\ee
\textbf{D6 with $\mathbf{B_{12}, B_{34}, B_{56}}$}
\bea
\label{D6BBB}
ds^{2}_{str} &=& f_{6}^{-1/2}[-dx^{2}_{0} + h_{1}(dx^{2}_{1}+dx^{2}_{2}) + h_{2}(dx_{3}^{2} + dx_{4}^{2}) + h_{3}(dx_{5}^{2} + dx_{6}^{2})] \nn &+& f_{6}^{1/2} (dr^{2} + r^{2} d \Omega_{2}^{2}), \nn
 B_{12} &=& \frac{s_1}{c_1} f_{6}^{-1} h_1, 
 \mbox{\hspace{5mm}} 
 B_{34} = \frac{s_2}{c_2} f_{6}^{-1} h_2, 
 \mbox{\hspace{5mm}} 
 B_{56} = \frac{s_3}{c_3} f_{6}^{-1} h_3, 
 \mbox{\hspace{5mm}}  f_{6} = 1+ \frac{Q_{6}}{r}, 
\nn
e^{ \phi} &=& f_{6}^{-3/4} (h_1 h_2 h_{3})^{1/2},  \mbox{\hspace{5mm}} h_{i}^{-1} = s_{i}^{2} f_{6}^{-1} + c_{i}^{2}, 
\eea
with the following fluxes
\be
\label{D6fluxes}
\begin{array}{ll}
 F_{r0} = s_1 s_2 s_3 \partial_{r} f_{6}^{-1},  &  
F_{r012} = -c_1 s_2 s_3 h_{1} \partial_{r} f_{6}^{-1} , \\ 
F_{r034} = -s_1 c_2 s_3 h_{2} \partial_{r} f_{6}^{-1} , & 
F_{r056} = -s_1 s_2 c_3 h_{3} \partial_{r} f_{6}^{-1},  \\  
F_{r03456} = -s_1 c_2 c_3 h_{2} h_{3} \partial_{r} f_{6}^{-1} ,  
&
F_{r01256} = -c_1 s_2 c_3  h_1 h_3 \partial_{r} f_{6}^{-1} ,  \\
F_{r01234} = -c_1 c_2 s_3 h_1 h_2 \partial_{r} f_{6}^{-1}, & 
 F_{r0123456} = -c_1 c_2 c_3 h_1 h_2 h_3 \partial_{r} f_{6}^{-1},  
\end{array} 
\ee
The projector is
\be
\label{D6BBBproj}
e^{\alpha_{1} \G_{12} \G_{11}} e^{\alpha_{2} \G_{34} \G_{11}} e^{\alpha_{3} \G_{56} \G_{11}} \G_{0123456} \e = \e, 
\ee
where 
\be
\label{D6framea}
\begin{array}{cc} \cos \alpha_{i} = c_{i} h^{1/2}_{i}, & \sin \alpha_{i} = s_{i} f_{6}^{-1/2} h_{i}^{1/2}\\ \end{array}.    
\ee

\section{D0-probes in D2, D4, D6 $B$-field backgrounds}
Having discussed the backgrounds with $B$-fields in the last section, we will consider the introduction a D0-probe, and its effect on the preserved supersymmetry. We initially consider string theory scattering amplitudes as a first approximation, before including the back-reaction of the D6 by performing a DBI probe calculation to determine the potentials seen by such probes. We confirm that these calculations overlap in the large distance limit. From \cite{Burgess:2003mm}, we know that D0-probes see an attractive, a flat and an repulsive potential for pure D2, D4 and D6-brane backgrounds respectively. In this section we see how the introduction of $B$-fields changes this analysis. 
 
\subsection{Kappa symmetry analysis}
Here we establish what to expect by examining a supergravity projector for a D0-brane 
\be
\G_{11} \G_{0} \e = \e, 
\ee
and considering its compatibility with the 1/2 BPS projectors from the last section. We will work in orthonormal frame where $B = \tan \alpha$, and will via this analysis, rederive the supersymmetry conditions.   

Introducing the D0-projector into the D2 with $B$-field background, means ensuring that the matrix $\G_{11} \G_{0} $ commutes with (\ref{projD2}). As $\G_{11} \G_{0}$ anti-commutes with the pure D2-projector $\G_{012}$, this is only possible in the limit that $\sin \theta \rightarrow 0$, or alternatively in the infinite $B$-field limit. In this limit, the final configuration again recovers half the supersymmetry. 

For the D4-background, by examining the projector again, we find the condition for supersymmetry
\be
\label{D4cond}
\sin (\alpha_{1} \pm \alpha_{2}) = 0, 
\ee   
where we have allowed for a choice of sign in the projector (\ref{projD4}), while imposing the D0 projector $\G_{11} \G_{0} \e = \e$ and the D4 projector $\G_{11} \G_{01234}$. This constraint above essentially removes the D2-projectors leaving the mutually commuting D0 and D4-projectors, making the final configuration 1/4 BPS. In terms of the $B$-fields it just allows (anti-)self-dual $B$-fields. 

Finally for the D6-background, we see that the D0-projector and D6-projector anti-commute. They can only be reconciled if we orchestrate the $B$-fields, so that we only impose D0 and D4 projectors (or alternatively, D2 and D6-projectors which are manifest in later solutions)
\be
\begin{array}{cccc} \G_{11} \G_{0}, & \G_{11} \G_{01234}, & \G_{11} \G_{01256} , & \G_{11} \G_{03456}\end{array},
\ee 
in the presence of the constraint
\be
\label{D6cond}
\cos(\alpha_{1} \pm \alpha_{2} \pm \alpha_{3}) = 0. 
\ee
This configuration is 1/8 BPS.  

\subsection{Scattering amplitudes}
In this section we calculate the force between static Dp-branes in string theory in the presence of $B$-fields. By considering the usual cylinder vacuum amplitude \cite{Polchinski}, we can weigh the attraction from the graviton, dilaton and $B$-fields with the repulsion due to the RR tensor. We simply quote the results with the details being removed to the appendix. These amplitudes we later compare with the DBI probe results in the large $R$ limit. 

Initially, we consider D0-brane located at a finite distance $R$ from D2, D4 and D6-branes with $B$-fields. For a D2-brane (stretched along directions $x_0, x_1$ and $x_2$) with a magnetic field $B_{12} = b$ on its worldvolume, the amplitude of the interaction is given by 
\be
{\mathcal A} \propto  T_{2}^2 g^2_s \ \frac{1}{\cos\theta}(1-\sin \theta)^2 G_7(R^2).
\ee
Here $b = \tan \theta$, and see that as the $B$-field increases the attraction between the  branes diminishes, until the limit $b \rightarrow \infty$, where there is no force. 

We next consider a D4-brane with $B$-fields interacting with a D0-brane.  The most general $B$-field in this case has four non-zero components $B_{12}=-B_{21}= b_1$ and $B_{34}=-B_{43}=  b_2$. Following similar analysis to above, the amplitude becomes
\be
{\mathcal A} \propto T_{4} T_0 g^2_s \ \frac{2-\cos2\theta_1-\cos2\theta_2-4\sin\theta_1\sin\theta_2 }{\cos \theta_1 \cos \theta_2 }
G_{5}(R^2),
\ee
where $b_{1} = \tan\theta_1$ and $b_2 = \tan\theta_2$. This amplitude vanishes when 
\be 
2-\cos2\theta_1-\cos2\theta_2-4\sin\theta_1\sin\theta_2 =0,
\ee 
which gives $\sin\theta_1=\sin\theta_2$ or equivalently $\theta_1=\theta_2$ corresponding to (anti-)self-dual $B$-fields. 

Finally we move onto a D6-brane with $B$-field interacting with a D0-brane.  The most general $B$-field in this case has six non-zero components $B_{12}=-B_{21}=  b_1,  B_{34}=-B_{43}=  b_2$ and $B_{56}=-B_{65}=  b_3$.  The same analysis gives the amplitude to be
\be
{\mathcal A} \propto T_{6} T_0 g^2_s \ \frac{1-\cos2\theta_1-\cos2\theta_2-\cos2\theta_3+4\sin\theta_1\sin\theta_2\sin\theta_3 }{6 \cos \theta_1 \cos \theta_2 \cos \theta_3}
G_{3}(R^2),
\ee
where $b_1 = \tan\theta_1, b_2 = \tan\theta_2$ and $b_3 = \tan\theta_3$. This amplitude vanishes when 
\be 
\label{nof}
1-\cos2\theta_1-\cos2\theta_2-\cos2\theta_3+4\sin\theta_1\sin\theta_2\sin\theta_3 =0
\ee 
This happens for $\theta_1+\theta_2+\theta_3= \pi/2 (\mbox{mod}\; 2\pi)$ or equivalently for $b_1b_2+b_2b_3+b_1 b_3 =1$. 

\subsection{DBI probe analysis} 
For the backgrounds introduced in section 2 we will consider D0-brane DBI probes. The action comprises of a Born-Infeld and a Wess-Zumino term,
\bea
S &=& S_{BI} + S_{WZ}, \nn
S &=& - T_{0} \int d \tau e^{-\phi} \sqrt{- \mathcal{P}[G+B]_{\tau \tau}} -q T_{0} \int \tilde{C}^{(1)}, 
\eea 
where $T_{0}$ is the tension of the probe. The value of $q$ depends on whether the probe is a brane ($+1$) or an anti-brane ($-1$), and $\mathcal{P}[G+B]$ denotes the pull-back of the background fields to the worldvolume of the D0-brane. $\tilde{C}^{(1)}$ refers to the induced D0-charge resulting from turning on $B$-fields in the presence of Dp-branes. In what follows we will make use of static-gauge. Similar analysis for D0-Dp without $B$-fields may be found in \cite{Burgess:2003mm}, which we follow. 

For the Dp with $B$-field backgrounds, the BI term takes the form:
\be
\label{tiltsbi}
S^{p}_{BI} = -m \int d \tau A_{p} \sqrt{1-f(\dot{r}^{2} + r^{2} \dot{\phi}^{2})},
\ee
where $A_{p}$ depends on Dp-brane $B$-field background we are probing:
\bea
A_{2} &=& f^{-1/2} h^{-1/2}, \nn
A_{4} &=& (h_{1} h_{2})^{-1/2}, \nn
A_{6} &=& f^{1/2} ( h_{1} h_{2} h_{3})^{-1/2}.
\eea
Here $m$ is just the tension of the D0 i.e. $T_{0} = m$. 

The WZ for $p=2,4,6$ may be read off from the supergravity solutions introduced earlier. 
We then proceed by deriving the canonical momenta $p_{i} = \partial L/ \partial q^{i}$ and the Hamiltonian.
The Hamiltonian is a monotonically increasing function of both $\dot{r}, \dot{\phi}$, so we set these terms to zero to find the potential. The potential $V$ derived from the Hamiltonian $H=m V$ then takes the simple form
\be
\label{derivedpot}
V \equiv A_{p} + q \tilde{C}^{(1)}.
\ee

We can now proceed case by case. We will be interested in analysing the potentials as the $B$-fields vary. In the case of D2 and D4 it is possible to tune $B$, by completing squares, such that $V$ is a constant and there is a no force (BPS condition). For D6, this was not possible but we plotted the potential and noted the minimum. 

For D2, the D0-probe sees a potential that gradually flattens as the $B$-field is increased until the potential becomes a constant. From
\be
V = f_{2}^{-1/2} \left( s^{2} f_{2}^{-1} + c^{2} \right)^{1/2} + q s f_{2}^{-1},
\ee
we see that only the choice $c=0$, $q=-1$, will make $V$ constant. This agrees with the earlier $\kappa$-symmetry analysis where we noted that in the limit of infinite $B$-field on the D2, the D2-charge is dissolved and the probe will only see D0-charge. In this limit there is no force. Similar features are seen for the later potentials, so from now on we confine ourselves to finite $B$-fields. 

For zero $B$-fields there is no force between D0 and D4. The addition of $B$-fields makes the potential attractive unless the $B_{12} = B_{34}$. To see this we complete squares so that $V$ maybe written
\bea
V &=& \biggl(s_1^{2} s_{2}^{2} f^{-2} + c_1^{2} c_2^{2} + \frac{f^{-1}}{2} \left( (c_{1} s_2 + s_1 c_2)^{2} + (c_{1} s_2 - s_1 c_2)^{2} \right)\biggr)^{1/2} \nn &-& q s_1 s_2 f^{-1}.
\eea
By confining ourselves to the first quadrant i.e. $c_i > 0, s_i >0$, we see that imposing 
\bea
s_1/c_1 &=& s_2/ c_2, \nn B_{12} &=& B_{34},  
\eea
leads to a constant potential $V = c^{2}$ if $q=1$. In orthonormal frame this above self-dual condition on the $B$-fields agrees with the earlier $\kappa$-symmetry (\ref{D4cond}). For this condition on the $B$-fields, the induced D2-charge does not attract the D0-probe and it sees only the source D4-charge and the induced D0-charge via the $B$-fields. Neither of these exert any force on the probe. 

For D6, the potential starts off repulsive in the absence of $B$-fields. As one increases the $B$-fields, there are two cases to consider. For $q <0$, the potential is repulsive. However for $q>0$, as the $B$-fields are increased beyond a certain value, the repulsion is overcome and the potential forms a bound state - Fig. 1. For the critical $B$-field value, this bound state is at infinity, but as the $B$-fields are increased further, the location of the bound approaches $r=0$. We determined the minimum of the potential as a function of coordinates $\theta_1, \theta_2 $ and $\theta_3$ and found that it was located at 
\bea
\label{rmin}
r &=& - \frac{\cos \theta_1  \cos \theta_2  \cos \theta_3 Q_{6}}{\cos(\theta_1 + \theta_2 + \theta_3)}, \\ 
\label{rminB}
&=& \frac{Q_{6}}{b_{1} b_{2} + b_{1} b_{3} + b_{2} b_{3} - 1} . 
\eea
In moving between the angles of (\ref{rmin}) and the $B$-fields of (\ref{rminB}), we have used $B_{2i \;2i-1}|_{\infty} = b_i = \tan \theta_{i}$. We have a lower bound on the existence of a supersymmetric D0-D6 system in terms of asymptotic $B$-fields: 
\be 
\label{Bbound}
b_{1} b_{2} + b_{1} b_{3} + b_{2} b_{3} 
\geq 1 . 
\ee
We see here that the $B$-fields have to be large enough to overcome the repulsion. 
The above location of the minimum may seem quite strange until it is repackaged in terms of orthonormal frame angles $\alpha_{i}$ (\ref{D6framea}), where it becomes (\ref{D6cond}). In other words, the D0 probe knows about Witten's supersymmetry conditions. This seems like a surprising result as we have come upon it in a rather circuitous manner.

The finite separation from the D6 in its supersymmetric configuration is evidence in higher dimensions that supersymmetric D0-D6 will be multi-centered \cite{Denef:2007vg}, where when (\ref{Bbound}) is saturated, one finds a marginal stability wall. This all rings well with the work of Denef and Moore in four-dimensions.

\begin{figure}
\label{D6D0fig}
 \begin{center}
  \includegraphics[scale=.7]{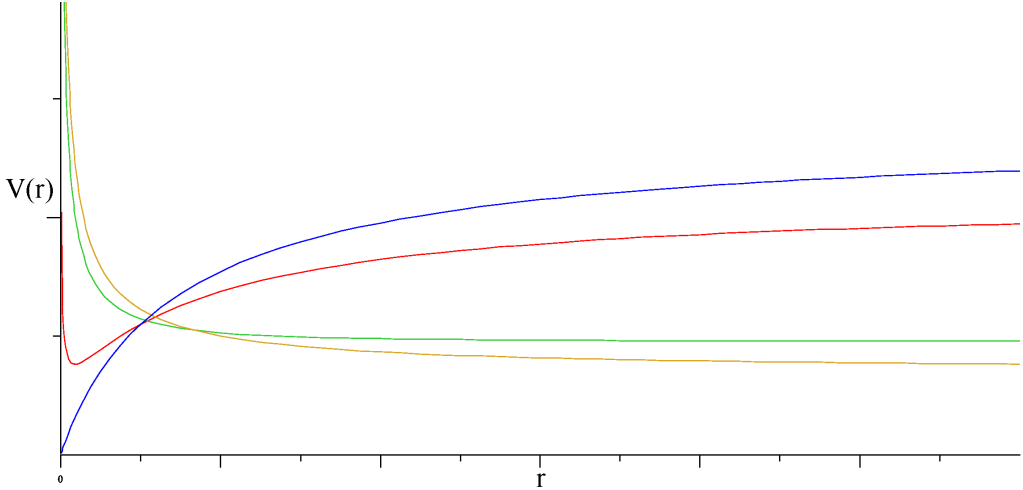}
 \end{center}
\caption{Potential $V(r)$ seen by D0 probe in D6 BBB background as the $B$-fields are increased. In ascending order one sees: a repulsive potential ($b=0$); a flat potential at infinity for the critical value $b_{crit}$; for $b > b_{crit}$ this bound state moves inwards; eventually in the large $B$-field limit the D0 is attracted.}
\end{figure}

In beautiful work, Denef, Moore and collaborators describe composite BPS bound states compactified on CY$_{3}$ from ten to four-dimensions \cite{Denef:2000nb,Denef:2007vg}. From \cite{Denef:2007vg}, the separation between a composite state of charges $\G_{1}$ and $\G_{2}$ is given by
\be
r = \frac{\langle \G_{1},\G_{2} \rangle}{2}\frac{|Z_1+Z_2|}{\mbox{Im}(Z_1\bar{Z}_{2})},
\ee
where $\langle \G,\Delta \rangle$ is an intersection product on $H^{3}(CY_{3},\mathbb{C})$.

For this system the holographic central charges are simply: 
\be
Z_{1} = p^{0} \tau^{1} \tau^{2} \tau^{3}, \quad
Z_{2} = -q_{0},  
\ee
where $p^{0}$ and $q_{0}$ are the D6 and D0-charge respectively and $b_{I} = \mbox{Re} \tau $ is the $I^{th}$ $B$-field and $a_{I} = \mbox{Im} \tau_{I}$ is the area of $I^{th}$ $T^{2}$ of $T^{6}$. By taking the probe approximation $p^{0} = Q_{6} c_1 c_2 c_3 >> q_{0}$ (\ref{D6fluxes}) limit for finite $b_{I}, a_{I}$, after correctly normalising the asymptotic behaviour, one finds that the distance between D0 and D6 is (\ref{rminB}),
with the pole being at the same point of moduli space. 

Just one more comment: The above potentials all harbour information about the geometry of the D6, while the string scattering amplitudes are all performed at the worldvolume of the brane. As a result, the gravity effects of the latter are largely overlooked, and to compare with the potentials seen by the D0-probes in the presence of Dp-brane, we must look at the large $R$ limit. We expand the DBI potentials (\ref{derivedpot}) in $R$
\be
\label{expandedV}
V(R) = V_{0} + V_{1}(B)/R^{7-p} +  \cdots, 
\ee 
where we use $V_{1}(B)$ to remind us that the potential is a function of the $B$-fields. The force, $-d V/dR$, which we may directly compare with the amplitude is then  
\be
F = V_{1}(B)/R^{7-p-1}. 
\ee
So, when $V_{1}(B) = 0$, we get a no-force condition for the $B$-fields in terms of $\theta_{i}$. Bearing in mind $q = \pm 1$, we summarise the results in the following:
\be
\begin{array}{cc} \mbox{Dp-brane} & \mbox{Condition on angles} \\
D2 & \sin \theta_1 = -q, \\
D4 & \sin \theta_1 = q \sin \theta_2, \\
D6 & \sin \theta_1 = \pm \cos (\theta_2 + q \theta_3).
\end{array}
\ee
As anticipated, we recover the string theory results. The large distance limit reconciles these two probe approaches.

\section{Supersymmetric D0-D6 solution}

The supergravity solution for a large class of three-charge supertube, black hole and black ring solutions in five-dimensions were found several years ago. For a decent review, we recommend \cite{Bena:2007kg}. These all allow uplifts on $T^{6}$ to M-theory and preserve at least 1/8 BPS. From our earlier $\kappa$-symmetry analysis in section three, we have seen that the desired D0-D6 solution with $B$-fields will preserve the same amount of supersymmetry. In this section, we identify that solution from the larger class. In particular, we identify the correct charges and investigate the conditions imposed on the solution by demanding it to be free of closed-timelike-curves (CTCs). 
\subsection{General solution}

For an eleven dimensional metric of the form 
\bea 
\label{mthmetric}
ds^2 &=& -(Z_1Z_2Z_3)^{-2/3} (d t +k)^2 + (Z_1Z_2Z_3)^{1/3} (ds_{B}^2) \cr\cr &+& (Z_1^{-2}Z_2Z_3)^{1/3}(dx_1^2+dx_2^2) + (Z_1Z_2^{-2}Z_3)^{1/3}(dx_3^2+dx_4^2) \cr\cr &+& (Z_1Z_2Z_3^{-2})^{1/3}(dx_5^2+dx_{6}^2), 
\eea 
with a one-form
\be
k = \mu (d z + \vec{A} \cdot d \vec{x}) + \omega, 
\ee
and a four-dimensional Gibbons-Hawking base metric $ds_{B}^2$:
\bea 
 ds_B^2 &=& V^{-1} (dz + p \cos \theta d \phi)^2 + V (dr^{2} + r^2 d \Omega_{2}^{2}), 
\label{eqA} \nn 
V &=& 1 + p/r, 
\eea 
the BPS conditions are satisfied if $Z_{I}$ and $\mu$ take the following form,
\bea 
Z_I &=& \frac{1}{2} C_{IJK} V^{-1} K^J K^K + L_I, \nn
\mu &=& \frac{1}{6}C_{IJK}\frac{K^IK^JK^K}{V^2} + \frac{1}{2V}K^IL_I +M, 
\eea
and $\omega $ solves the equation
\be
\label{eqomega}
\vec{\nabla} \times \vec{\omega} = V \vec{\nabla} M - M \vec{\nabla} V + \frac{1}{2}(K^I \vec{\nabla} L_I - L_I \vec{\nabla} K^I).
\ee
Here $K^{I}, L_{I}, M$ and $V$ are all harmonic functions allowing multiple centres and in the case of $T^6$, we have $C_{IJK} = |\epsilon_{IJK}|$. 

The M-theory three-form potential $A^{(3)}$ is given by
\be
A^{(3)} = A^{1} \wedge dx^{12} + A^{2} \wedge dx^{34} + A^{3} \wedge dx^{56}, 
\ee 
where the one-form potentials $A^{I}$, may be expressed thus:
\be
\label{flux11}
A^{I} = \frac{K^{I}}{V} (d z + p \cos \theta d \phi) + \vec{\beta}^{I} \cdot d \vec{x} -\frac{1}{Z_{I}}(d t +k),
\ee
with $\beta$ denoting the solution to 
\be
\vec{\nabla} \times \vec{\beta}^{I} = - \vec{\nabla} K^{I}. 
\ee

Having skimmed over the general form of the solution in M-theory, we now reduce to IIA so that we can make contact with the earlier single-centred D6 $B$-field solution.  
The ten-dimensional solution is then
\bea 
\label{IIAmetric}
ds^2 &=& -f^{-1/2}(d t + \omega)^{2} + f^{1/2} V^{-1} Z_{I}^{-1} (dx_{2I-1}^{2} + dx_{2}^{2}) + f^{1/2} (dr^2+r^2d\Omega_2^2), \nn
e^{\phi} &=& f^{3/4} V^{-3/2} (Z_1 Z_2 Z_3)^{-1/2},\nn
\label{pots}
C^{(1)} &=& - \frac{\mu V^{2}}{f} (d t + \omega) + p \cos \theta d \phi, \nn   
C^{(3)} &=&  \left[-\frac{1}{Z_{I}} (d t + \omega) + \beta^{I}\right] \wedge dx^{2I-1} \wedge dx^{2I} +  p \cos \theta  d \phi \wedge B^{(2)}, \nn  
B^{(2)} &=& \left[ \frac{K^{I}}{V} - \frac{\mu}{Z_{I}} \right] dx^{2I-1} \wedge dx^{2I}, 
\eea
with summation over $I$ and 
\be
f = Z_1 Z_2 Z_3 V - \mu^2 V^{2}.
\ee
To proceed, we need to establish a connection between the coefficients appearing in the harmonic functions
and the asymptotic D6, D4, D2 and D0 charges.  

From \cite{Bena:2007kg}, we see that the eight functions of the general solution  $V, K^{I}, L_{I}, M$ maybe identified with the eight independent parameters in the $\mathbf{56}$ of the $E_{7(7)}$ duality group in four dimensions: 
\be
\label{identity}
p^{0} = -V, \quad p^{I} = K^{I}, \quad q_{I} = L_{I}, \quad q_{0} = -2 M.
\ee
With these identifications, the quartic invariant $I_{4}$, 
\cite{Kallosh:1996uy}
 takes the form
\bea
I_{4} &=& q_{0} p^{1} p^{2} p^{3} - p^{0} q_1 q_2 q_3 - \left( p^{0} q_{0} + p^{I} q_{I} \right)^{2} + 4 \sum_{I < J} p^{I} q_{I} p^{J} q_{J}, \nn
\label{fCTC}
&=& -M^{2} V^{2} - \frac{1}{3} M C_{IJK} K^I K^J K^K - M V K^{I} L_{I} - \frac{1}{4} (K^{I} L_{I})^{2} \nn &+& 
\frac{1}{6} V C^{IJK} L_I L_J L_K + \frac{1}{4} C^{IJK} C_{IMN} L_{J} L_{K} K^{M} K^{N}. 
\eea
Although the entropy does not depend on the sign of $I_{4}$, it is important as it separates BPS black hole solutions ($I_{4} > 0$) from non-BPS solutions ($I_{4} < 0$). The non-BPS D0-D6 solutions with $B$-fields were analysed in \cite{Gimon:2007mh}.

We note that the harmonic functions $K^{I},L_{I}$ and $M$ correspond to D4, D2 and D0-charge respectively. These are in addition to the D6 charge. For D6, D6 with $B$-fields and D0-D6 with $B$-fields, neither D2 nor D4 charges appear, so we will henceforth set $K^{I}$ and $L_{I}$ to be constants 
\be
K^{I} = k^{I}_{0}, \quad L_{I} = l_{I0}. 
\ee
This choice will be validated later when we calculate the charges. 

\subsection{Solution constraints}
As the metric may be shown to be regular even when $V=0$ \cite{Bena:2007kg}, we only need examine the presence of CTCs. We primarily concern ourselves with ensuring the metric has the correct signature asymptotic signature $\eta_{\mu \nu}$ and with eliminating of Dirac-Misner strings. 

The first condition may be imposed by demanding that 
the inequality
\be
\label{CTCcond}
\quad f r^{2} \sin^{2} \theta - \omega_{\phi}^{2} > 0, 
\ee  
holds everywhere. 

For later purposes, in analysing the second constraint from Dirac-Misner strings, we consider a two-centre solution of finitely separated D0-charge $m_2$ from D6-D4-D2-D0, 
\bea
\label{ourcharges}
K^{I}  &=& k^{I}_0 + \frac{k^{I}_1}{r}, \nn
L_{I}  &=& l_{I0} + \frac{l_{I1}}{r}, \nn
M &=& m_0 + \frac{m_1}{r} + \frac{m_2}{\Sigma}, 
\eea
where $\Sigma = \sqrt{r^{2} + R^{2} - 2 R r \cos \theta}$. This solution corresponds to a solution located on the $z$-axis of $\mathbb{R}^{3}$ at $z=0$ and $z = R$. The azimuthal angle is given by $\theta$.  

When solving for $\omega$ (\ref{eqomega}), one encounters  three kinds of terms on the right hand side
\be
\vec{\nabla} \frac{1}{r}, \quad \vec{\nabla} \frac{1}{\Sigma}, \quad \frac{1}{r} \vec{\nabla} \frac{1}{\Sigma}-\frac{1}{\Sigma} \vec{\nabla} \frac{1}{r}.
\ee
These respectively admit the following solutions for $\omega_{\phi}$:
\be
\cos\theta, \quad \frac{r \cos \theta - R}{\Sigma}, \quad \frac{r-R \cos \theta}{R \Sigma},
\ee
with the general solution being a linear combination of these with the addition of a constant $\kappa$. 
With the above choice of harmonic functions, $\omega_{\phi}$ is
\bea
\omega_{\phi} &=& [m_1 -m_0 p + \frac{1}{2} (k^{I}_0 l_{I1} -l_{I0} k^{I}_1)] \cos \theta 
+ m_2 
\left( \frac{r \cos \theta -R}{\Sigma} \right) \nn
&+& p m_2  
\left( \frac{r -R \cos \theta}{R \Sigma} \right) + \kappa, 
\eea
Requiring that Dirac-Misner strings vanish on the $z$-axis corresponds to demanding $\omega_{\phi}(\theta = 0, \pi)  = 0$. In terms of the above coefficients, this condition can be met if 
\be
\label{dmstr}
[m_1 -m_0 p + \frac{1}{2} (k^{I}_0 l_{I1} -l_{I0} k^{I}_1)] = -m_2 = \frac{m_2 p }{R}  = -\kappa. 
\ee
The final expression for $\omega_{\phi}$ then becomes
\be
\omega_{\phi} = m_2\left[1 - \frac{(r+R) }{\Sigma} \right](1-\cos \theta). 
\ee
Note here that the vanishing of Dirac-Misner strings imposes the asymptotic flatness condition, $\omega_{\phi} \rightarrow 0$ as $r \rightarrow \infty$, for free. 

\subsection{D6 solutions}

The simplest example we consider is single-centred D6. From (\ref{identity}), the absence of D0-charge means that $m_{1} = 0$. It also leads to CTCs and Dirac-Misner strings, so it should be set to zero. For similar reasons $m_{0} = 0$.  At this point, only 
\be
\label{D6brane}
k^{I}_{0} = 0,  \quad l_{I0} = 1, 
\ee    
will lead to a solution with asymptotic metric $\eta_{\mu \nu} $ and no $B$-fields present.  

Next we can consider adding $B$-fields to the D6. Again the absence of CTC requires $m_0 = 0$.  If we define $b_{I}$ to be the asymptotic value of the $B$-field at infinity from (\ref{IIAmetric}) and denote the string coupling constant by $g_s=e^{\Phi} |_{\infty}$, we have

\bea 
b_{I} &=&-\frac{\sum_{J} k^{J}_{0} l_{J0} - 2 k^{I}_0 l_{I0} }{(|\epsilon_{IJK}| k^{J}_0 k^{K} _{0}+ 2 l_{I0})} \\ g_s^{4/3} &=& \frac{ \sum_{I \neq J} k^{I}_0 l_{I0} k^{J}_{0} l_{J0}  + \frac{2}{3} |\e_{IJK} | l_{I0} l_{J0} l_{ K0} - \sum_{I} (k^{I}_{0}l_{I0})^{2} }{\prod_{I} (2 l_{I0} + |\e_{IJK}| k^J_0 k^K_0 )^{2/3}}\;.
\eea 
These can be used to find $k^{I}_0$ and $l_{I0}$ as follows
\be 
k^{I}_0 = \frac{(\sum_{J \neq I}  b_J )(b_{I}^2+g_s^{4/3})}{\sum_{J < K} b_J b_K-g_s^{4/3}}\; ,\;\;\;\;\;\;\; l_{I0} = -\frac{\prod_{J \neq I} (b_{J}^2+g_s^{4/3})}{\sum_{J < K} b_J b_K -g_s^{4/3}}
\ee

For simplicity, we take $g_s=1$ henceforth. Finally to get the flat metric $\eta_{\mu\nu}$ asymptotically, we need to rescale coordinates $r$ and $t$ in (\ref{IIAmetric}) by\footnote{Since we take $g_s=1$ we do not need to rescale $T^6$ coordinates. In general we need to rescale it by factor $g_s^{3/4}$. }
\be 
\tilde{t} = f_{\infty}^{-1/4} t \; , \;\;\;\;\;\;\; \tilde{r} = f_{\infty}^{1/4} r
\ee 
where we have denoted the asymptotic value of $f$ at infinity by $f_{\infty}$ which is given by
\be
f_{\infty} = \frac{\prod_{I} (b_{I}^2+1)^2}{(\sum_{I < J}b_I b_J -1)^4}
\ee

Using these relationships one may then plough ahead and calculate the asymptotic charges. Taking into account the rescaling one finds,
\bea
q_{0} &=& \frac{1}{\kappa^{2}} \int_{T^{6} \times S^{2}} \star d C^{(1)} = 
-\frac{4 \pi}{\kappa^2} Vol(T^{6}) p b_{1} b_2 b_3, \nn 
q_{I} &=& \frac{1}{\kappa^{2}} \int_{T^{4} \times S^{2}} \star \left( d C^{(3)} - H \wedge C^{(1)} \right) = 
 \frac{2 \pi}{\kappa^{2}} Vol(T_{I}^{4}) |\e_{IJK}| p b_{J} b_{K}, \nn 
p^{I} &=& \frac{1}{\kappa^{2}} \int_{T^{2} \times S^{2}} d C^{(3)} = 
-\frac{4 \pi}{\kappa^{2}} Vol(T^{2}_{I} ) p b_{I}, \nn 
p^{0} &=& \frac{1}{\kappa^{2}} \int_{T^{2} \times S^{2}} d C^{(1)} = 
-\frac{4 \pi}{\kappa^{2}} p. 
\eea 
The four-dimensional mass may also be calculated using the rescaled metric
\be
4 G_{4} M = p \prod_{I} (1+b_{I}^2)^{1/2}. 
\ee
These charges agree with those computable using the earlier metric (\ref{D6BBB}) and fluxes (\ref{D6fluxes}) corresponding to the 1/2-BPS D6 with $B$-fields solution. 

Before leaving this example, there is one final remark. As $\omega_{\phi} = 0$, we only require  $f > 0$ everywhere for this solution to be CTC-free. Expanding $f$, one sees that it is positive if, $p \prod_{I} l_{I0}  >0$, or alternatively, if
\be
p(1-\sum_{I<J} b_I b_J) > 0. 
\ee
Now most of the work has been done. We simply have to introduce a D0-charge to the mix. As seen above, $m_{1}$ is necessarily zero to avoid CTCs. So the presence of CTCs rules out the introduction of non-induced D0-charge on top of the D6-brane. In other words, there is no single-centred \textit{supersymmetric} D0-D6 solution. The only way to add a D0-charge then seems to be to turn on $m_{2}$, which corresponds to the addition of D0-charge at a finite distance $\tilde{R}$. Here we are using the rescaled metric. 

Analysis of the vanishing of Dirac-Misner strings (\ref{dmstr}) in the rescaled metric imposes the following constraints
\bea
\label{cond1}
m_{2} &=& m_{0} p, \\
\label{cond2}
\tilde{R} &=& -f_{\infty}^{1/4} p. 
\eea
The first condition (\ref{cond1}) here is also required to satisfy (\ref{CTCcond}), so it is consistent. The second sets $p < 0$, which as mentioned before, causes no problems for regularity. 

We now again solve for $k_{0}$ and $l_{0}$  in terms of the new asymptotic $B$-field
\be
b^{I} =-\frac{p(\sum_{J}  k^{J}_{0} l_{J0} - 2 k^{I}_0 l_{I0}) +m_{2}}{p(|\epsilon_{IJK}| k^{J}_0 k^{K} _{0}+ 2 l_{I0})},
\ee and find
\be
\label{genl0}
k^{I}_0 = \frac{(p \sum_{J \neq I}   b_J )(b_{I}^2+1) + 2 m_2}{p(\sum_{J < K} b_J b_K-1)},  \quad  l_{I0} = -\frac{p \prod_{J \neq I} (b_{J}^2+1) + 2 m_2 \sum_{I \neq J} b_I}{p(\sum_{J < K} b_J b_K -1)}.
\ee
Therefore, the distance $\tilde{R}$ is given by
\be 
\tilde{R} = \frac{( p^2 \prod_{I} (1+b_{I}^2) +4 m_{2} \left(m_{2} + p (\sum_{I} b_{I} - \prod_{I} b_{I})\right) )^{1/2}}   {\sum_{I < J} b_I b_J -1}, 
\ee 
To compare this with the DBI calculation, we simply take the $m_2 \rightarrow 0 $ limit. In this limit
\bea
\tilde{R} &=& \frac{|p| \prod_{I} (b_I^2+1)^{1/2} }{ \sum_{I < J} b_I b_J -1}, \nn
&=& \frac{Q_{6}}{b_1b_2 +b_2 b_3 +b_1 b_3-1}, 
\eea
where in the last line, we recover the same result as the DBI.  

This solution is again CTC-free if $p \prod_{I} l_{I0} > 0$, where $l_{I0}$ are given above (\ref{genl0}). 

Despite the dependence of the $B$-field on the additional D0-charge $m_{2}$, one can recalculate the charges. After a little bit of algebra, one finds that the charges with three independent $B$-fields are
\bea
q_{0} &=&  \frac{4 \pi}{\kappa^{2}} Vol(T^{6}) (-p b_1 b_2 b_3 + 2 m_{2}), \nn 
q_{I} &=&  |\epsilon_{IJK}| \frac{2 \pi}{\kappa^{2}} Vol(T_{I}^{4}) p b_{J} b_{K}, \nn 
p^{I} &=& -\frac{4 \pi}{\kappa^{2}} Vol(T_{I}^{2}) p b_{I}, \nn 
p^{0} &=& -\frac{4 \pi}{\kappa^{2}} p. 
\eea
The ADM mass and angular momentum may be expressed 
\bea
4 G_{4} M &=& \left( p^2 \prod_{I} (1+b_{I}^2) +4 m_{2} \left(m_{2} + p (\sum_{I} b_{I} - \prod_{I} b_{I})\right) \right)^{1/2}, \\
\label{angmom}
{\mathcal J} &=& \frac{m_2|p|}{2G_4}.
\eea
\subsection{Black hole generalisation} 
The motivation so far has been to see how D0 interacts with D6 in the presence of $B$-fields. We have noted the presence of three regimes dependent on the $B$-fields. An immediate generalisation is to consider D6 with extra charges and $B$-fields and to once again look at how the forces balance themselves out in a supersymmetric setting. Recall that we expect the potential seen by D0 to have an attractive contribution from D2 charges, a repulsive contrbution from D6, with D0 and D4 playing the role of onlookers. In principle, via scattering and DBI probe calculations, one can get better acquainted with this system by ignoring various degrees of back-reaction. 

With the solution constructed in the previous section, it is an easy task to consider D0 in the presence of D6-D4-D2-D0 with $B$-fields system. Refering the reader to (\ref{ourcharges}), we are considering $m_2$ D0-charge at one centre, while turning on $k^{I}_{1}, l_{I1}$ and $m_1$ on the D6. The charges for this system take the rather simple form:
\bea
q_{0} &=&  \frac{4 \pi}{\kappa^{2}} Vol(T^{6}) (-p b_1 b_2 b_3 + \frac{1}{2!} |\e_{IJK}| k^{I}_{1} b_{J} b_{K} + l_{I0} b_I  + 2 (m_1 + m_{2})), \nn 
q_{I} &=&  \frac{4 \pi}{\kappa^{2}} Vol(T_{I}^{4}) (\frac{1}{2!} | \e_{IJK}| p b_{J} b_{K}  - |\e_{IJK}| k^{J}_{1} b_{K} -l_{I1}), \nn 
p^{I} &=& \frac{4 \pi}{\kappa^{2}} Vol(T_{I}^{2}) (-p b_{I} + k^{I}_{1}), \nn 
p^{0} &=& -\frac{4 \pi}{\kappa^{2}} p, 
\eea
where we consider sums over contracted indices. One can clearly see how the $B$-fields induce lower dimensional Dp-brane charges. But, in general, we don't expect this more general two-centred configuration to preserve supersymmetry i.e. we expect to run into CTCs. 

However, we have explicitly checked that for a range of the parameters there exists a CTC-free supersymmetric solution when only D2-charges are present i.e. $k^{I}_{1} = m_1 = 0$. Here we present the case where all the $L_{I}$ and $b_{I}$ are equal, with any generalisation being again immediate. 
The expression for the asymptotic $B$-fields is then
\be
b =-\frac{p k_{0} l_{0}  + 2m_{2} + 3k_0 l_1 }{2 p( k^{2}_0 + l_{0})}.
\ee
We use (\ref{dmstr}) to eliminate $m_0$, which along with the finite separation guarantees there are no Dirac-Misner strings. The expressions for $k_0$ and $l_0$ in terms of $p, l_1, m_2$ and $b$ are
\be
k_0 = \frac{2b p(b^2+1) + 2 m_2}{p(3 b^2 -1) -3 l_1},  \quad  l_{0} = -\frac{(p + 3 l_1) (b^2+1) + 4 b  m_2}{p(3 b^2-1) - 3 l_1 }.
\ee
The distance between the two centres then becomes  
\be
\tilde{R} = - f_{\infty}^{1/4} p, 
\ee
where 
\be
f_{\infty} =  \frac{ p^2 (1+b^2)^3 + 4 m_2 (m_2 + p b(3 - b^2)) + (9 l_1^2 + 6 l_1 p (1-b^2))(1+b^2) + 12 b l_1 m_2}{(3 p b^2 -p - 3 l_1)^{4}} . 
\ee
The angular momentum of this solution is unchanged from (\ref{angmom}). This is not surprising as we haven't added a D4 magnetic partner for the D2 at the position of the D0-charge $m_2$. The mass may be expressed as
\be
4 G_{4} M = f_{\infty}^{1/4} (3 p b^2 -p - 3 l_1). 
\ee
The solution will be CTC-free again if (\ref{CTCcond}) is satisfied everywhere. 

Reducing to four-dimensions, there is a horizon at $\tilde{r}=0$. The Beckenstein-Hawking entropy is given by 
\be
\mathcal{S}_{BH} = \sqrt{l_1^3 p}{G_{4}}.
\ee

\section{Discussion}
In this work we investigated the physics of the supersymmetric D0-D6 system. Our study culminates in writing down explicitly a 1/8 BPS solution. In the process of this work, we also glance over simpler Dp-brane systems with $B$-fields. By probing the D6 with $B$-fields background with a D0, the result solidifies our understanding of the dependence of the D0-D6 solution on $B$-fields. We see that there is a wall of marginal stability and a two-centred supersymmetric D0-D6 only exists if the asymptotic $B$-fields are sufficiently large. Once this value is exceeded, the separation distance decreases with increasing $B$-field. 

In constructing the final solution, we also had to make use of one extra ingredient. From electromagnetics, we expect a system which carries both electric and magnetic charges to generate angular momentum, so our final solution necessarily carries angular momentum. In terms of the existing five-dimensional black hole and black ring literature, we see how that absence of Dirac-Misner strings and the correct signature of the metric (no CTCs) dictate the rest of the story: they rule out a single-centred D0-D6 and determine the distance of separation between the sources as a function of the asymptotic $B$-fields. Although this solution is not a black hole, we generalise the solution by adding extra charges to the D6, so that the D6 develops a horizon. 

It would be interesting to consider the D0-dynamics from the perspective
of the noncommutative Yang-Mills theory derived from the D6-branes with
nonzero $B$-fields. Our gravity contruction implies that the BPS object
should carry nonzero R-charge, which is somewhat different from what see
in the field theory.

Another open avenue is to consider generalisations of the above D0-D6 solution to D0-D6-D4-D2-D0 with supersymmetry. One can ask how the addition of more centres helps preserve supersymmetry. We can also consider charges at the location of the D0, which should lead to black ring solutions \cite{Emparan}. Within these generalisations there will be black objects allowing microstate descriptions. 
\section*{Acknowledgements}
We are grateful to I. Bena, D. Jatkar, F. Denef, R. Emparan, J. Gauntlett, K. Hosomichi, N. Kim, A. Sen, S. Sheikh-Jabbari, E. Witten and P. Yi for discussion and correspondence. HY would also like to express his gratitude to L. Alvarez-Gaume, N. Brambilla, C. Gomez, S. Ferrara and E. Rabinovici for
support during an excellent workshop on Black Holes at CERN. KML is supported in part by KRF-2006-C00008 and KOSEF SRC Program
through CQUeST at Sogang University. KPY is grateful to KIAS and HRI for hospitality and was supported by the Korea Science and Engineering Foundation (KOSEF) grant funded by the Korea government(MEST) through the Center for Quantum Spacetime(CQUeST) of Sogang University with grant number R11 - 2005 - 021.
\appendix
\section{Conventions}
\subsection{T-duality}
The action of T-duality on massless NS-NS sector fields $G_{mn}, B_{mn}$, and the dilaton $\phi$ is well known. In search of consistent conventions, we choose to adopt the conventions of Hassan \cite{Hassan:1999bv} wholesale. In the case of the NS fields these are: 
\be
\begin{array}{ll}
\tilde{G}_{zz} = 1/G_{zz}, & e^{2 \tilde{\phi}} = e^{2 \phi} / G_{zz}, \\
\tilde{G}_{\mu \nu} = G_{\mu \nu} -(G_{z \mu} G_{ z \nu } - B_{z \mu } B_{z \nu } )/G_{zz}, & \tilde{G}_{ z \mu} = -B_{z \mu} /G_{zz},   \\
\tilde{B}_{\mu \nu} = B_{\mu \nu} -(G_{z \mu} B_{z \nu} - B_{z \mu} G_{z \nu} )/G_{zz}, & \tilde{B}_{z \mu} = -G_{z \mu} /G_{zz}.   \\
\end{array}
\ee
Here $z$ denotes the Killing coordinate in which direction we T-dualise, while $\mu,\nu$ denote  coordinates other than $z$. 
The RR fields which are independent of $z$ transform under T-duality as 
\bea
\tilde{C}^{(n)}_{z\nu_2...\nu_n} &=& a\left[ C^{(n-1)}_{\nu_2...\nu_n} -(n-1)(G_{z[\nu_2}C^{(n-1)}_{z\nu_3...\nu_n]})/G_{zz} \right], \nn
\tilde{C}^{(n)}_{\nu_1 \nu_2...\nu_n} &=& a C^{(n+1)}_{z \nu_1 \nu_2...\nu_n} -nB_{z[\nu_1}\tilde{C}^{(n)}_{z\nu_2...\nu_n]}.
\eea
Throughout this paper, we will adopt the $a=+1$ convention. 

\subsection{D=11,10 Supergravities}
We will follow the conventions of \cite{Gauntlett:2003wb} in using a (-,+,+,...) space signature with $\e_{012...\sharp} = +1$. The inner product of a $q$-form with a $p$-form is 
\be
\alpha \lrcorner \beta = (1/q!) \alpha^{b_1...b_{q}} \beta_{b_1...b_q a_1...a_{p-q}},
\ee
and the Hodge dual of a $q$form in $D$ dimensions is defined by 
\be
\label{hodge}
\star \alpha_{b_1...b_q} = (1/q!) \e_{b_1...b_{D-q}}^{\phantom {abcdefgh}  a_1...a_q} \alpha_{a_1...a_q}. 
\ee

In $D=11$, imposing supersymmetry requires that the variation of the gravitino $\Psi_{M}$ be zero:
\be
\label{D11KSE}
\d \Psi_{M} = \nabla_{M} \e + \frac{1}{12} \left[ \G_{M} \mathbf{G}^{(4)} - 3 \mathbf{G}^{(4)}_{M} \right] \e = 0,
\ee 
where we define the contractions in bold via
\bea
\mathbf{A}^{(n)} &=& \frac{1}{n!} A_{i_1...i_n} \G^{i_1...i_n}, \nn
\mathbf{B}_{m}^{(n)} &=& \frac{1}{(n-1)!} B_{m i_2...i_n} \G^{i_2...i_n}. 
\eea
Here 
\be
\nabla_{M} \equiv \partial_{M} + \frac{1}{4} \omega_{MAB} \G^{AB},
\ee
where the spin connection $\omega_{MAB}$ (in any dimension) is calculable from the vielbein
\bea
\omega_{MAB} &=& \frac{1}{2} (-c_{MAB} + c_{ABM} -c_{BMA}), \nn
c_{MN}^{\phantom {AB} A} &=& 2 \partial_{[N} E^{A}_{{M} ]}.
\eea

$D=11$ supergravity metrics are related to IIA metrics in $D=10$ via the reduction ansatz
\be
\label{redansatz}
ds^{2}_{M} = \mbox{exp} (-2\phi/3) ds^{2}_{IIA} + \mbox{exp} (4 \phi/3) \left(d x_{11} + C^{(1)}\right)^{2}.  
\ee
In performing this reduction, in addition to the $D=10$ IIA metric we also introduce a scalar field $\phi$ (dilaton) and a one-form potential $C^{(1)}$. The three-form $A^{(3)}$ and the field strength $G^{(4)} = d A^{(3)}$ in $D=11$ are then decomposed as 
\bea
A^{(3)} &=& C^{(3)} + B \wedge d x_{11}, \nn
G^{(4)} &=& F^{(4)} + H \wedge \left(d x_{11} + C^{(1)} \right) , 
\eea
where 
\bea
H &=& dB, \nn
F^{(4)} &=& d C^{(3)} - H \wedge C^{(1)}. 
\eea
Taking into account the warp-factor $e^{-2 \phi/3}$ in (\ref{redansatz}), we see that 
\be
\mathbf{G}^{(4)} = e^{4 \phi/3} \mathbf{F}^{(4)} + e^{\phi/3} \mathbf{H}^{(3)} \G_{11}. 
\ee
The warp-factor will also produce extra terms via the spin connection when we take the above reduction ansatz and place it in (\ref{D11KSE}). If we then make the following redefintions:
\bea
\lambda &=& 3 e^{-\phi/6} \G_{11} \Psi_{11}, \nn
\psi_{m} &=& e^{-\phi/6} (\Psi_{m} + \frac{1}{2} \G_{m} \G_{11} \Psi_{11}), \nn
\e &=& e^{-\phi/6} \tilde{e},
\eea
we obtain the Killing spinor equations of IIA
\bea
\label{dilatino}
\d \lambda  &=& \left[ \partial_{a} \phi \G^{a} - \frac{1}{2} \mathbf{H}^{(3)} \G_{11} - \frac{3}{4} e^{\phi} \mathbf{F}^{(2)} \G_{11} +  \frac{1}{4 } e^{\phi} \mathbf{F}^{(4)} \right] \tilde{\e}, \\
\label{gravitino}
\d \psi_{m} &=& \left[ D_{m} - \frac{1}{4} \mathbf{H}_{m}^{(2)} \G_{11} - \frac{1}{8} e^{\phi} \mathbf{F}^{(2)} \G_{m} \G_{11} +  \frac{1}{8} e^{\phi} \mathbf{F}^{(4)} \G_{m} \right] \tilde{\e}. 
\eea

\subsection{Equations of Motion for IIA}
We begin with the bosonic form of the supergravity analysis from Polchinski \cite{Polchinski}. The action may be written:
\bea
S_{IIA} &=& S_{NS} + S_{R} + S_{CS}, \nn
&=& \frac{1}{2 \kappa^{2}_{10}}  \int \biggl[  e^{-2 \phi} \left( R \star \mathbf{1} + 4 d \phi \wedge (\star d \phi) - \frac{1}{2} H \wedge (\star H) \right) \nn
&+& \frac{1}{2} F^{(2)} \wedge (\star F^{(2)}) + \frac{1}{2} \tilde{F}^{(4)} \wedge (\star \tilde{F}^{(4)} ) 
+ \frac{1}{2} B \wedge F^{(4)} \wedge F^{(4)} \biggr], 
\eea 
where \footnote{We have flipped the sign of $B$ in this action so that our definitions of the gauge invariant four-form $\tilde{F}^{(4)}$ coincide with Polchinski. }
\be
\tilde{F}^{(4)} = F^{(4)} - H \wedge C^{(1)}.  
\ee
We also note that we have defined the volume form such that 
\be
dx^{a0} \wedge dx^{a1} \wedge \cdots \wedge dx^{a9} = \sqrt{-g} \e^{a0a1...a9} dx^{0} \wedge dx^{1} \wedge \cdots \wedge dx^{9},  
\ee
and our Hodge-duality conventions are unchanged from before (\ref{hodge}). 
Varying this action with respect to $B$, $C^{(1)}$ and $C^{(3)}$ respectively we get the following flux equations of motion:
\bea
\label{IIAeom}
0 &=& d(e^{-2 \phi} \star H) + d(C^{(1)} \wedge \star \tilde{F}^{(4)}) + \frac{1}{2} F^{(4)} \wedge F^{(4)}, \nn
0 &=&   d (\star F^{(2)} ) + H \wedge \star \tilde{F}^{(4)}, \nn
0 &=& d (\star \tilde{F}^{(4)} ) +H \wedge F^{(4)}. 
\eea
\section{D2 $B$-field Killing spinors}
We list the gravitino variations for D2 with $B$-field (\ref{D2B}) here. As mentioned in the main text, the relationship between $\e$ and $\tilde{\e}$ is given by (\ref{etildee}). 
\bea
\d \psi^{(B)}_{0} &=& \frac{1}{8} f^{3/4} \partial_{r} f^{-1} \G^{r0} \e - \frac{1}{8} f^{3/4} \partial_{r} f^{-1} \G^{r 12} e^{-\alpha \G_{12} \G_{11}} \e, \nn
&=& e^{\alpha/2 \G_{12} \G_{11}} \left[\frac{1}{8} f^{3/4} \partial_{r} f^{-1} \G^{r0} \right] (1 - \G_{012} ) \tilde{\e}, \nn
\d \psi^{(B)}_{1} &=&  e^{3\alpha/2 \G_{12} \G_{11} } \left[ - \frac{1}{8} f^{3/4} \partial_{r} f^{-1} \G^{r1} \right] (1- \G_{012} ) \tilde{\e}, \nn
\d \psi^{(B)}_{2} &=&  e^{3\alpha/2 \G_{12} \G_{11} } \left[ - \frac{1}{8} f^{3/4} \partial_{r} f^{-1} \G^{r2} \right] (1- \G_{012} ) \tilde{\e}, \nn
\d \psi^{(B)}_{r} &=& f^{-1/4} \partial_{r} \e - \frac{1}{2} f^{-1/4} \partial_{r} \alpha  \G_{12} \G_{11} \e - \frac{1}{8} f^{3/4} \partial_{r} f^{-1} \e ,\nn
&=& e^{\alpha/2 \G_{12} \G_{11} } \left[ f^{-1/4} \partial_{r} - \frac{1}{8} f^{3/4} \partial_{r} f^{-1} \right] \tilde{\e}. 
\eea
We may also check the variation of the gravitino in one of the external $\theta$ directions on the transverse sphere, getting
\be
e^{\alpha_{1}/2 \G_{12} \G_{11}} \left[ \frac{f^{-1/4}}{r} (\partial_{\theta} - \frac{1}{2} \G^{r \theta}) + \frac{1}{8} f^{3/4} \partial_{r} f^{-1} \G^{r \theta} (1 - \G_{012} ) \right] \tilde{\e}. 
\ee
We will ignore the external variations of the gravitino in all subsequent analysis, confident that these variations are zero. In each case the variations will simply give us information about $\e$ i.e in the case of D2 we get
\be
\e = e^{\theta/2 \G_{r \theta}}e^{\phi/2 \G_{\theta \phi}} \eta,
\ee
where $\eta $ is a constant spinor satisfying the projector $e^{\alpha \G_{12} \G_{11}} \eta = \eta$. 
\section{Dp-Dp' bound state with B-field}

We consider the interaction between a Dp-brane which are stretched along directions $x^0, \cdots x^p$ and located at $x^{i} =0, i=p+1 \cdots 9$ and a Dp'-brane stretched along directions $x^0, \cdots x^{p'}$ and located at $x^j = Y^j, j=p'+1 \cdots 9$. The open strings which are stretched between these D-branes are described by following boundary conditions

\be
\sigma=0 \left\{  \begin{array}{cc}
X^{\mu}=0 & \mu=p+1,..,9 \\ 
\partial_{\sigma}X^{\mu}=0  &  \mu=0,..,p
\end{array}\right.
\ee
\be
\sigma=\pi \left\{  \begin{array}{cc}
X^{\mu}=Y^{\mu} & \mu=p'+1,..,9 \\ 
\partial_{\sigma}X^{\mu}=0  &  \mu=0,..,p'
\end{array}\right.
\ee
Boundary conditions on world-sheet fermions will be given by supersymmetry transformation. We find

\be
\begin{array}{cc}
 X^{\mu}=p^{\mu}\tau +\sum_{n \in Z}\frac{1}{n} \alpha_n^{\mu} e^{-in\tau} \cos n\sigma &  \mu= 0,...,p' \\
  =\sum_{r \in Z+1/2} \frac{1}{r} \alpha_r^{\mu} e^{-ir\tau} \sin r\sigma &  \mu=p'+1,...,p  \\      
 \ \ \ \ \  =Y^{\mu} \frac{\sigma}{\pi}+\sum_{n \in Z} \frac{1}{n} \alpha_{n}^{\mu}e^{-in\tau} \sin n\sigma &  \mu=p+1,...,9, 
\end{array}
\ee
where for the {\bf{R}}-sector $\psi_{\pm}^{\mu}$:

\be
\left\{ \baa{cc}
\hspace{-18mm} \psi^{\mu}_+=\sum_{n \in Z} d^{\mu}_n e^{-in(\tau+\sigma)} \;\;\; \hspace{7mm}
\psi^{\mu}_-=\sum_{n \in Z} d^{\mu}_n e^{-in(\tau-\sigma)} \;\;\;\mu=0,...,p'  \\
\hspace{-8mm} \psi^{\mu}_+=\sum_{n \in Z} d^{\mu}_n e^{-in(\tau+\sigma)} \;\;\; \hspace{7mm}
\psi^{\mu}_-=-\sum_{n \in Z} d^{\mu}_n e^{-in(\tau-\sigma)} \;\;\;\mu=p+1,...,9 \\
\psi^{\mu}_+=\sum_{r \in Z+1/2} d^{\mu}_r e^{-ir(\tau+\sigma)} \;\;\; 
\psi^{\mu}_-=-\sum_{r \in Z+1/2} d^{\mu}_r e^{-ir(\tau-\sigma)} \;\;\;\mu=p'+1,...,p 
\eaa \right\}
\ee
while for the {\bf{NS}}-sector 
\be
\left\{ \baa{cc}
\psi^{\mu}_+=\sum_{r \in Z+1/2} b^{\mu}_r e^{-ir(\tau+\sigma)} \;\;\; 
\psi^{\mu}_-=\sum_{r \in Z+1/2} b^{\mu}_n e^{-ir(\tau-\sigma)} \;\;\;\mu=0,...,p'  \\
\psi^{\mu}_+=\sum_{r \in Z+1/2} b^{\mu}_r e^{-ir(\tau+\sigma)} \;\;\; 
\psi^{\mu}_-=-\sum_{r \in Z+1/2} b^{\mu}_n e^{-ir(\tau-\sigma)} \;\;\;\mu=p+1,...,9 \\
\psi^{\mu}_+=\sum_{n \in Z} b^{\mu}_n e^{-in(\tau+\sigma)} \;\;\; 
\psi^{\mu}_-=-\sum_{n \in Z} b^{\mu}_n e^{-in(\tau-\sigma)} \;\;\;\mu=p'+1,...,p 
\eaa \right.
\ee
and the quantization condition in terms of mode expansions are given by 
\be 
[\alpha_r^{\mu},\alpha_s^{\nu}]=\delta_{r+s} \delta^{\mu \nu}, 
\ee
\be 
\{d_r^{\mu},d_s^{\nu}\}=\delta_{r+s} \delta^{\mu \nu},
\ee
\be 
\{b_n^{\mu},b_m^{\nu}\}=\delta_{n+m} \delta^{\mu \nu}. 
\ee
Therefore the mass spectrum for {\bf{NS}}-sector  is given by 
\be
\alpha'M^2=\frac{Y^2}{4\pi^2\alpha'}+N-(\frac{1}{2}-\frac{\Delta}{8}), \;\;\;\;\, 
\Delta=p-p',
\ee
\be
N=\sum_{n>0} \alpha_{-n}.\alpha_{n}+\sum_{r>0} \alpha_{-r}.\alpha_{r}+
\sum_{r>0} rb_{-r}.b_{r} + \sum_{n>0} nb_{-n}.b_{n}, 
\ee
and for the {\bf{R}}-sector it is given by 
\be
\alpha'M^2=\frac{Y^2}{4\pi^2\alpha'}+N, \;\;\;\;\,\Delta=p-p',
\ee
\be
N=\sum_{n>0}^{8-\Delta} \alpha_{-n}.\alpha_{n}+\sum_{r>0}^{\Delta} \alpha_{-r}.\alpha_{r}+
\sum_{n>0}^{8-\Delta} nd_{-n}.d_{n} + \sum_{r>0}^{\Delta} rd_{-r}.d_{r}. 
\ee
D-branes can interact by exchanging closed strings. This can be expressed in terms of open string loops. As a result, the amplitude is given by
\be 
{\mathcal A}=\int \frac{dt}{2t} \sum_{i,p}e^{-2\pi\alpha' t (p^2+M_i^2)}
\ee 
After a bit algebra this amplitude can be written as\footnote{We need to consider appropriate GSO projection.}
\be
{\mathcal A}=2V_{p'+1} \int \frac{dt}{2t}(8\pi^2\alpha't)^{-(p'+1)/2}e^{-\frac{Y^2t}{2\pi^2\alpha'}} 
 (\bf{ NS-R}), 
\ee
where $\bf{ NS}$ and $\bf{ R}$ are given by 
\be
{\bf NS} = 2^{\Delta/2-1}\ q^{-1+\frac{\Delta}{4}}
\biggl(\prod{\frac{(1-q^{2n})}{(1-q^{2n-1})}}\biggr) ^{\Delta}
\biggl(\prod{\frac{(1+q^{2n})}{(1+q^{2n-1})}}\biggr) ^{\Delta}
\biggl(\prod{\frac{(1-q^{2n})}{(1+q^{2n-1})}}\biggr) ^{-8}
\ee
\be
{\bf R}= 2^{3-\Delta/2}
\biggl(\prod{\frac{(1-q^{2n})}{(1-q^{2n-1})}}\biggr) ^{\Delta}
\biggl(\prod{\frac{(1+q^{2n-1})}{(1+q^{2n})}}\biggr) ^{\Delta}
\biggl(\prod{\frac{(1-q^{2n})}{(1+q^{2n})}}\biggr)   ^{-8}
\ee
Now if we restrict ourselves to massless closed string exchange (small $t$ limit) we get
\be
{\mathcal A} = V_{p'+1}(4\pi^2\alpha')^{3-\frac{p+p'}{2}}\ {(2-\Delta/2)}\pi 
G_{9-p}(Y^2),
\ee
where $G_{9-p}$ is massless Green function in $9-p$-dimensions. As an example, one may consider D0-D2 where
\be
{\mathcal A} = V(4\pi^2\alpha')^{2}\pi G_{7}(Y^2),
\ee
the positive amplitude implies there is an attractive force between D0 and D2.

Having recapped the procedure, we now shift focus and consider two parallel Dp-branes and turn on a $B$-fields on the worldvolume of one of these branes. We start by examining the case where the $B$-field has just two non-zero components ($B_{p-1\ p}=-B_{p\ p-1}=2\pi \alpha' b$). The boundary conditions are given by 
\bea
&& \sigma=0 \left\{  \begin{array}{cc}
\partial_{\sigma}X^{\mu}=0 \;\;\;\;\  \mu=0,1,...,p \\
 X^{\mu}=0 \;\;\;\;\ \mu=p+1,...,9 . 
\end{array}\right.
\cr\cr
&& \sigma=\pi \left\{  \begin{array}{cc}
\partial_{\sigma}X^{\mu}=0 \;\;\;\;\  \mu=0,1,...,p-2 \\
\hspace{-16mm}\partial_{\sigma}X^{p-1}+b \partial_{\tau}X^p=0 \\
\hspace{-16mm}\partial_{\sigma}X^{p}-b \partial_{\tau}X^{p-1}=0 \\
 \hspace{-2mm} X^{\mu}=Y^{\mu} \;\;\;\;\ \mu=p+1,...,9.  
\end{array}\right.
\eea
After a little algebra, the amplitude may be determined to be
\be
{\mathcal A} \propto V_{p} T_{p}^2 g^2_s \ \frac{1}{\sin \theta}(1-\cos \theta)^2 
G_{9-p}(Y^2)
\ee
where, $B=\tan \theta$. 

Some things to note here: if we switch off the $B$-field $\theta = 0$, $\mathcal{A} = 0$ and there is no force between the branes. For $\theta > 0$, we get attraction ($\mathcal{A} > 0$). So we see the attracting influence of the $B$-fields on what was an initially BPS configuration of parallel branes.

\end{document}